\documentclass[doublecol]{epl2} 

\usepackage{amssymb}
\usepackage{amsmath}
\usepackage{color}
\usepackage{units}

\newcommand*{\TRANS}{^{\mkern-1.5mu\mathsf{T}}}
\newcommand*{\VEC}[1]{\boldsymbol{#1}}
\newcommand*{\TENSOR}[1]{\mathsf{#1}}
\newcommand*{\OP}[1]{{\cal{#1}}}
\newcommand*{\kB}{k_{\mathrm{B}}}

\newcommand*{\EPS}{\varepsilon}

\DeclareMathOperator{\Tr}{Tr}

\title{Diffusion of a Brownian ellipsoid in a force field}

\author{Erik Aurell\inst{1,2}
\and Stefano Bo\inst{3}
\and Marcelo Dias\inst{3,4}
\and Ralf Eichhorn\inst{3}\thanks{E-mail: \email{eichhorn@nordita.org}}
\and Raffaele Marino\inst{3}}
\shortauthor{E. Aurell \etal}

\institute{                    
\inst{1}
Dept. of Computational Biology and Center for Quantum Materials,
KTH -- Royal Institute of Technology,  AlbaNova University Center, SE-106 91~Stockholm, Sweden
\\
\inst{2}
Depts. Information and Computer Science and Applied Physics, Aalto University, Espoo, Finland
\\
\inst{3}
Nordita, Royal Institute of Technology and Stockholm University,
Roslagstullsbacken 23, SE-106 91 Stockholm, Sweden
\\
\inst{4}
Aalto Science Institute, School of Science, Aalto University, FI-02150 Espoo, Finland
}


\abstract{We calculate the effective long-term convective velocity and
dispersive motion of an ellipsoidal Brownian particle in three dimensions
when it is subjected to
a constant external force. This long-term motion results
as a ``net'' average behavior
from the particle rotation and translation
on short time scales.
Accordingly, we apply a systematic multi-scale technique
to derive the effective equations of motion valid on long times.
We verify our theoretical results by comparing them to numerical simulations.}

\begin{document}

\maketitle


\section{Introduction}
The Brownian motion \cite{duplantier05,mazo02} of
small particles suspended in an (aqueous) solvent,
driven by the erratic impacts from the solvent molecules,
is an ubiquitous phenomenon below micrometer length scales.
Its theoretical foundations have been studied
for over 100 years, with huge impact on the natural
sciences in general and on physics in particular
\cite{frey05,hanggi05}.
Still, there are many puzzles of surprisingly
fundamental nature which are not yet fully resolved.
An important example is the effect of hydrodynamic coupling
between the translation and rotation of particles with
arbitrary, non-spherical shape,
like colloidal particles, colloidal clusters, DNA, proteins,
nanotubes etc., on their overall diffusive behavior.
This problem has recently attracted considerable interest
\cite{han06,han09,gonzalez10,cichocki12,cichocki15,grima07,ribrault07},
presumably spurred by the developments in
single particle tracking techniques which are able to
record both, position and orientation, with high precision
\cite{han06,han09,fung13,chakrabarty13,chakrabarty14}.

Brownian motion of non-spherical particles is characterized
by a crossover from short-term anisotropic diffusion,
dominated by the initial particle orientation, to
effective ``net'' diffusion on very long times.
For free Brownian motion without external forces,
this behavior has been studied in some detail and is quite well understood
\cite{han06,gonzalez10,cichocki12,cichocki15}.
The situation is different, however, if the diffusive
motion is driven by an external force.
The only theoretical studies in that direction,
that we are aware of, 
are a recent analysis of driven Brownian motion of
an asymmetrical particle in the plane \cite{grima07},
and a work by Brenner published in 1981 and little noted in the physical literature~\cite{brenner81}.
Having in mind the settling of a dilute suspension of small ellipsoidal
particles due to gravitation,
Brenner studied this problem under the term ``sedimentation dispersion''.
However, for calculating this dispersion, i.e.\ the
\emph{effective long-term diffusion coefficient},
from the full description of particle translation and orientation,
Brenner used what he himself calls an ``ad-hoc approach'', and which
in fact is not explained in any detail in~\cite{brenner81}.


In the present paper we analyze the effective long-term motion
of a Brownian ellipsoidal particle in three dimensions,
using a systematic
multi-scale perturbation scheme. 
This scheme does not require an explicit parametrization
of rotations in three dimensions, but can rather be performed on a
relatively general and abstract level.
We are thus able to fully recover the pioneering results by Brenner
using an approach which over the last decades has become more
established, and which opens up the prospect for further generalizations.
We furthermore compare our results to numerical simulations
of the coupled equations of motion for translation
and rotation, as far as we know for the first time.

\section{Model}
We model the dynamics of the Brownian ellipsoid by
the force and torque balance relations
\begin{subequations}
\label{eq:balance}
\begin{eqnarray}
0 & = & -\gamma \dot{\VEC{x}} + \VEC{f} + \sqrt{2\kB T} \gamma^{1/2} \VEC{\xi}(t)
\label{eq:Ftrans}
\, , \\
0 & = & -\eta \VEC{\omega} + \sqrt{2\kB T} \eta^{1/2} \VEC{\zeta}(t) \, .
\label{eq:Frot}
\end{eqnarray}
\end{subequations}
The right-hand sides comprise all (non-inertial) forces and torques
which add up to zero, because
we neglect inertia effects \cite{purcell77,dusenbery11} (overdamped approximation).
The term $-\gamma \dot{\VEC{x}}$ represents the viscous friction force acting on
the particle center $\VEC{x}=(x_1,x_2,x_3)$, given by the 
particles' translational velocity $\dot{\VEC{x}}$ 
multiplying the friction tensor $\gamma$. All the externally applied forces
are collected in $\VEC{f}=(f_1,f_2,f_3)$.
Throughout this paper, we consider only \emph{constant} external force fields
$\VEC{f}$, independent of particle position and orientation.
The last term in \eqref{eq:Ftrans} models thermal fluctuation
by unbiased Gaussian white noise sources $\VEC{\xi}(t)=(\xi_1(t),\xi_2(t),\xi_3(t))$
with correlations $\langle \xi_i(t)\xi_j(t') \rangle = \delta_{ij}\delta(t-t')$.
The strength of these fluctuations is characterized
by the thermal energy $\kB T$ ($\kB$ is Boltzmann's
constant and $T$ the temperature) and the friction tensor.
Since $\gamma$ is symmetric and positive definite, its square root $\gamma^{1/2}$ is
uniquely defined, i.e.\ $\gamma^{1/2}\gamma^{1/2}=\gamma$.
The thermal environment is assumed to be homogeneous,
so that $T$ and $\gamma$ are constant in space.
In \eqref{eq:Frot}, all quantities represent the rotational counterparts
to the ones from \eqref{eq:Ftrans},
with the (positive definite and symmetric) rotational friction tensor $\eta$
(independent of position $\VEC{x}$),
its square root $\eta^{1/2}$, the angular velocity $\VEC{\omega}$,
and unbiased Gaussian noise sources $\VEC{\zeta}(t)$
which are independent of $\VEC{\xi}(t)$.
We restrict ourselves to the case without externally applied
torques so that only viscous friction and
thermal fluctuations contribute to the torque balance.

In principle, the dynamics of the Brownian ellipsoid is not
fully specified yet by eqs.~\eqref{eq:balance}.
While the force balance \eqref{eq:Ftrans} indeed represents
an equation of motion for
the translational Brownian movement of the center of the ellipsoid
(overdamped Langevin equation \cite{mazo02,snook07}),
the torque balance \eqref{eq:Frot} is 
a kinematic relation for the momentary angular velocity
induced by the acting torques. For a full description,
it has to be supplemented by a representation of
the particle orientation and its equation of motion.
As \eqref{eq:balance} is written in the laboratory frame,
the friction tensors $\gamma$ and $\eta$ then directly depend on the parameters
specifying the particle orientation,
i.e.\ the equations \eqref{eq:balance} are coupled.
Common representations of rotations in three dimensions
include Euler angles \cite{goldstein80}, quaternions \cite{coutsias04}
and unit vectors (``directors'')
attached to the particle \cite{brenner81,gonzalez10,marino16}.
It turns out, surprisingly, that the analysis
we are going to present in the following
can be performed without choosing a specific
representation of orientation, so that the description
provided by eqs.~\eqref{eq:balance} is sufficient for our purposes.

As already mentioned in the Introduction,
the driven diffusive motion of the ellipsoid 
described by \eqref{eq:balance} shows a crossover from short-term
anisotropic behavior, dictated by the particle's initial orientation,
to effective long-term convection and diffusion,
after the initial orientation has been ``forgotten''.
This crossover is characterized by the time it takes the
particle to diffusively perform a full rotation. We can
estimate it as \cite{ribrault07}
\begin{equation}
\label{eq:tau_c}
\tau_c = 1/\bar{Q} \, ,
\end{equation}
with $\bar{Q} = \Tr(\TENSOR{Q})/3$ and 
$\TENSOR{Q}=\kB T\eta^{-1}$ the rotation diffusion tensor.
Being the average of the eigenvalues of $\TENSOR{Q}$,
the choice of $\bar{Q}$ as a ``net'' rotational diffusion
is physically intuitive.
The length scale $l_c$ associated with $\tau_c$
is given by the typical distance the particle covers
by convection and diffusion during $\tau_c$,
\begin{equation}
l_c = \max \left\{ \tau_c |\VEC{f}|\Tr(\gamma^{-1})/3 \, , \; \sqrt{\tau_c \bar{D} } \right\} \, ,
\end{equation}
where we again used averaged friction and diffusion coefficients,
in particular $\bar{D} = \Tr(\TENSOR{D})/3 = \kB T \Tr(\gamma^{-1})/3$
with the translation diffusion tensor $\TENSOR{D}=\kB T \gamma^{-1}$.

On time and length scales $\tau_L$ and $L$,
much larger than $\tau_c$ and $l_c$,
the ellipsoid will perform an ``effective'' translational motion,
incorporating its rotational diffusion in an averaged way.
We can therefore define the small dimensionless parameter
\begin{equation}
\label{eq:eps}
\EPS = \tau_c/\tau_L \ll 1
\end{equation}
to quantify the separation of the small scales $\tau_c$ and $l_c$
from the large scales $\tau$ and $L$.
Note that the long scales do not directly appear in the
model \eqref{eq:balance},
but rather
are introduced by the question we ask: What is the effective
long-term dynamics of the ellipsoid after transients have
died out?
This is unlike other typical problems involving distinct
scales, where these scales are inherent to
the problem so that a small (or large) parameter
explicitly appears already in the equations of motion
\cite{pagitsas86,mazzino05,marino16}.
Nevertheless, also in the present case
the standard multi-scale or homogenization technique
\cite{vergassola97,bender99,pavliotis08}
can be applied as a systematic perturbation procedure
to derive the sought effective equations.

\section{Multi-scale analysis}
The starting point of the multi-scale analysis
is the (forward) Fokker-Planck equation for the probability density
$p$ that the ellipsoid has a certain position and orientation at
time $t$ \cite{mazo02,gardiner83,vankampen87}:
\begin{subequations}
\label{eq:FPLdagger}
\begin{equation}
\label{eq:FP}
\frac{\partial p}{\partial t} - \left( \OP{L}^{\dagger} + \OP{M}^{\dagger} \right) p = 0 \, ,
\end{equation}
with
\begin{equation}
\label{eq:Ldagger}
\OP{L}^\dagger =
-\frac{\partial}{\partial x_i} \left[ \left( \gamma^{-1}\VEC{f} \right)_i - \frac{\partial}{\partial x_j} {D}_{ij} \right] \, .
\end{equation}
\end{subequations}
Here, we use index notation and the summation convention
to sum over repeated indices;
$D_{ij}$ are the components of the diffusion tensor $\TENSOR{D}$.
The operator $\OP{M}^\dagger$ in \eqref{eq:FP}
represents the generator of
rotary diffusion and thus
involves the Laplace-operator in rotation space whose
specific form depends on the choice of parametrization
of orientation. 
It is clear that this Fokker-Planck
equation resolves the probability density on all
scales.
The essential step to disentangle small and large scales
is to explicitly
introduce them as \emph{independent} variables
and to presume that $p$ is a function of
all these variables.
Using the symbol $\VEC{\alpha}$ for collecting
the three parameters representing orientation
(without specifying them in any further detail),
we define
\begin{subequations}
\label{eq:scales}
\begin{equation}
\label{eq:xalphascales}
\tilde{\VEC{x}} = \EPS^0 \VEC{x}
\, , \quad
\VEC{X} = \EPS^{1} \VEC{x}
\, , \quad
\tilde{\VEC{\alpha}} = \EPS^0 \VEC{\alpha}
\, ,
\end{equation}
and
\begin{equation}
\label{eq:tscales}
\theta = \EPS^0 t
\, , \quad
\vartheta = \EPS^{1} t
\, , \quad
\tau = \EPS^{2} t
\, ,
\end{equation}
\end{subequations}
and require that
\begin{equation}
\label{eq:p}
p=p(\theta,\vartheta,\tau,\tilde{\VEC{x}},\VEC{X},\tilde{\VEC{\alpha}}) \, .
\end{equation}
With these definitions, $\VEC{X}$ is of order one only for very large $\VEC{x}$,
and thus represents the large scale, which we are interested in.
Likewise, the time variables $\vartheta$ and $\tau$ become of order one at
large times $t$, where we expect from their relative scaling with respect to $\VEC{X}$
that on $\vartheta$ scales long-term convective motion occurs
and on $\tau$ scales long-term diffusive motion.
The small scale variables $\tilde{\VEC{x}}$ and $\tilde{\VEC{\alpha}}$
essentially correspond to the original variables $\VEC{x}$ and $\VEC{\alpha}$,
but are restricted to small scales by imposing periodic
boundary conditions for $p$, eq.~\eqref{eq:p}, in
$\tilde{\VEC{x}}$ and $\tilde{\VEC{\alpha}}$.
The spatial periodicity is assumed to be $l_c$,
whereas the small rotational scales $\tilde{\VEC{\alpha}}$ are
obviously periodic by definition.
Relevant rotational motion occurs on small scales only.
Accordingly, there is no large scale rotational variable
defined in \eqref{eq:xalphascales},
and the long-term effective equations of motion
will not include rotational degrees of freedom explicitly.

As a consequence of \eqref{eq:scales} and \eqref{eq:p},
the time and spatial derivatives in \eqref{eq:FPLdagger} turn
into
\begin{equation}
\label{eq:derivatives}
\frac{\partial}{\partial t}
= \frac{\partial}{\partial\theta} + \EPS \frac{\partial}{\partial\vartheta} + \EPS^2 \frac{\partial}{\partial\tau}
\, , \quad
\frac{\partial}{\partial x_i}
= \frac{\partial}{\partial\tilde{x}_i} + \EPS \frac{\partial}{\partial X_i}
\, ,
\end{equation}
while the generator of rotational diffusion
$\OP{M}^\dagger$ 
remains unchanged, in particular it does not
involve any terms containing $\EPS$.
In view of the scale separation \eqref{eq:eps}
we can treat $\EPS$ as a small perturbative
parameter and expand $p$ in powers of $\EPS$,
\begin{equation}
\label{eq:pexpansion}
p = p^{(0)} + \EPS p^{(1)} + \EPS^2 p^{(2)} + \ldots
\, ,
\end{equation}
where all $p^{(i)}$ \emph{a priori} inherit the functional dependence
\eqref{eq:p} on the various variables.
In these variables, $p^{(0)}$ is normalized to one, while all other
$p^{(i)}$ with $i>0$ are normalized to zero.
Plugging \eqref{eq:derivatives} and \eqref{eq:pexpansion} into
\eqref{eq:FPLdagger}, and collecting terms of equal powers in
$\EPS$ in the resulting expression, we obtain a
hierarchy of inhomogeneous Fokker-Planck like equations of which
we list the first three (order $\EPS^0$, $\EPS^1$ and $\EPS^2$):
\begin{widetext}
\begin{subequations}
\label{eq:hierarchy}
\begin{eqnarray}
\frac{\partial p^{(0)}}{\partial\theta} - \left( \tilde{\OP{L}}^{\dagger} + \tilde{\OP{M}}^{\dagger} \right) p^{(0)} & = & 0
\label{eq:hierarchy0}
\\
\frac{\partial p^{(1)}}{\partial\theta} - \left( \tilde{\OP{L}}^{\dagger} + \tilde{\OP{M}}^{\dagger} \right) p^{(1)} & = &
- \frac{\partial p^{(0)}}{\partial\vartheta}
- \frac{\partial}{\partial X_i} v_i p^{(0)}
+ 2 \frac{\partial}{\partial \tilde{x}_i}\frac{\partial}{\partial X_j} {D}_{ij} p^{(0)}
\label{eq:hierarchy1}
\\
\frac{\partial p^{(2)}}{\partial\theta} - \left( \tilde{\OP{L}}^{\dagger} + \tilde{\OP{M}}^{\dagger} \right) p^{(2)} & = &
- \frac{\partial p^{(0)}}{\partial\tau} -\frac{\partial p^{(1)}}{\partial\vartheta}
- \frac{\partial}{\partial X_i} v_i p^{(1)}
+ \frac{\partial}{\partial X_i}\frac{\partial}{\partial X_j} {D}_{ij} p^{(0)}
+ 2 \frac{\partial}{\partial \tilde{x}_i}\frac{\partial}{\partial X_j} {D}_{ij} p^{(1)}
\label{eq:hierarchy2}
\end{eqnarray}
\end{subequations}
\end{widetext}
\begin{floatequation}
\mbox{\textit{see eqs.~\eqref{eq:hierarchy}}}.
\end{floatequation}
In \eqref{eq:hierarchy}, we introduce the 
velocity vector $\VEC{v} = \gamma^{-1}\VEC{f}$
and the tilde over the operators $\tilde{\OP{L}}^\dagger$ and
$\tilde{\OP{M}}^\dagger$ to
indicate that they act on the small scale variables
$\tilde{\VEC{x}}$ and $\tilde{\VEC{\alpha}}$, respectively.
Note that \eqref{eq:hierarchy0} is \emph{exactly the same}
equation as \eqref{eq:FP}, with the essential difference, however,
that $p^{(0)}$ obeys \emph{periodic} boundary conditions in the
variables $\tilde{\VEC{x}}$ and $\tilde{\VEC{\alpha}}$.

For finding the solutions of the equation hierarchy \eqref{eq:hierarchy}
we largely follow the standard procedure detailed, e.g., in \cite{pavliotis08}.
We are interested in solutions of \eqref{eq:hierarchy}
which are stationary
on small scales after short-term transients have died out.
Hence, the desired solutions do not depend on $\theta$ such that we
can set $\partial p^{(i)}/\partial\theta = 0$ for all $i$.
A further important observation is that $\tilde{\OP{M}}^\dagger$
and $\tilde{\OP{L}}^\dagger$ do not depend on the large scale variable
$\VEC{X}$. The solution to \eqref{eq:hierarchy0} is therefore given
by a product ansatz
\begin{equation}
\label{eq:p0}
p^{(0)}=w(\tilde{\VEC{x}},\tilde{\VEC{\alpha}})\rho^{(0)}(\vartheta,\tau,\VEC{X}) \, ,
\end{equation}
where $w$ has to be normalized over $\tilde{\VEC{x}}$ and $\tilde{\VEC{\alpha}}$,
and $\rho^{(0)}$ over $\VEC{X}$.
Exploiting that $\VEC{v}$ in $\tilde{\OP{L}}^\dagger$ is constant in space
and that dependencies on particle orientation enter only via $\gamma^{-1}$
(likewise in $\tilde{\OP{M}}^\dagger$ where the rotational diffusion
coefficient depends on orientation via $\eta^{-1}$),
we find $w$ to be uniform for all $\tilde{\VEC{x}}$
and $\tilde{\VEC{\alpha}}$, with a constant
value set by normalization: $w=1/(4\pi l_c^3)$.
This uniform distribution carries
a probability current $\VEC{v}w$ (see \eqref{eq:Ldagger}),
corresponding to an averaged particle velocity
\begin{equation}
\label{eq:V}
\VEC{V} = \int \mathrm{d}\tilde{\VEC{x}}\mathrm{d}\tilde{\VEC{\alpha}}\, \VEC{v}w
= \overline{\gamma^{-1}\VEC{f}} = \overline{\gamma^{-1}}\VEC{f} 
\, ,
\end{equation}
where the overbar denotes the average over
the uniform orientational distribution.

Solving eqs.\ \eqref{eq:hierarchy1} and \eqref{eq:hierarchy2} is a little
more involved because of the inhomogeneities on the right-hand sides.
For a non-trivial solution to exist, they have to fulfill a so-called
solvability condition \cite{pavliotis08}, stating that these inhomogeneities have
to be orthogonal to the null-space of the operator $\tilde{\OP{M}}+\tilde{\OP{L}}$
adjoint to $\tilde{\OP{M}}^\dagger+\tilde{\OP{L}}^\dagger$.
The nullspace of $\tilde{\OP{M}}+\tilde{\OP{L}}$ contains all constants
(in $\tilde{\VEC{x}}$ and $\tilde{\VEC{\alpha}}$), so that the solvability
condition for (\ref{eq:hierarchy1}) reads
$\int \mathrm{d}\tilde{\VEC{x}}\mathrm{d}\tilde{\VEC{\alpha}}
\left( - \frac{\partial p^{(0)}}{\partial\vartheta}
- \frac{\partial}{\partial X_i} v_i p^{(0)}
+ 2 \frac{\partial}{\partial \tilde{x}_i}\frac{\partial}{\partial X_j} {D}_{ij} p^{(0)} \right) = 0$.
Inserting our above result \eqref{eq:p0} for $p^{(0)}$, we find
\begin{equation}
\label{eq:FPconvective}
\frac{\partial\rho^{(0)}}{\partial\vartheta} + \frac{\partial}{\partial X_i} V_i \rho^{(0)} = 0 \, .
\end{equation}
With this equation (and using $p^{(0)}=w\rho^{(0)}$) we can
simplify \eqref{eq:hierarchy1} to,
\begin{equation}
\label{eq:FP_p1}
\left( \tilde{\OP{M}}^\dagger + \tilde{\OP{L}}^\dagger \right) p^{(1)}
= w(v_i-V_i)\frac{\partial\rho^{(0)}}{\partial X_i} \, .
\end{equation}
As $w(v_i-V_i)$ is a function of the small scale
variables while $\partial\rho^{(0)}/\partial X_i$
depends on large scales only, we can solve \eqref{eq:FP_p1}
again by a product ansatz.
We thus set
\begin{equation}
\label{eq:p1}
p^{(1)} = w \lambda_{ij}(\tilde{\VEC{x}},\tilde{\VEC{\alpha}})f_j
\,\frac{\partial\rho^{(0)}}{\partial X_i}(\vartheta,\tau,\VEC{X}) \, ,
\end{equation}
where the factor $w$ has been introduced for later convenience.
Likewise, also the specific expression $\lambda_{ij}f_j=(\lambda\VEC{f})_i$
for the unknown vector function of the small scales
involving an auxiliary tensor
$\lambda=\lambda(\tilde{\VEC{x}},\tilde{\VEC{\alpha}})$
proves to be very convenient.

The ansatz \eqref{eq:p1} solves \eqref{eq:FP_p1} provided that the
functions $\lambda_{ij}$ solve the auxiliary equation
$( \tilde{\OP{L}}^\dagger + \tilde{\OP{M}}^\dagger ) \lambda_{ij}f_j = v_i-V_i
=(\gamma^{-1}-\overline{\gamma^{-1}})_{ij} f_j$,
where we used the definition $\VEC{v}=\gamma^{-1}\VEC{f}$ and eq.~\eqref{eq:V}
for rewriting the right-hand side inhomogeneity
$v_i-V_i$ in the last step.
Observing that this inhomogeneity,
as well as all $v_i$ and $D_{ij}$ appearing in $\tilde{\OP{L}}^\dagger$,
are independent of $\tilde{\VEC{x}}$, we conclude
that the $\lambda_{ij}$ do not depend on $\tilde{\VEC{x}}$ either,
so that $( \tilde{\OP{L}}^\dagger + \tilde{\OP{M}}^\dagger ) \lambda_{ij}f_j = \tilde{\OP{M}}^\dagger \lambda_{ij}f_j$.
As $\lambda_{ij}$ is supposed to be
a solution for any (constant) $\VEC{f}$,
the above auxiliary equation thus
simplifies to the tensor equation
\begin{equation}
\label{eq:FP_lambda}
\tilde{\OP{M}}^\dagger \lambda_{ij} = (\gamma^{-1})_{ij}-(\overline{\gamma^{-1}})_{ij} \, .
\end{equation}
In addition, all $\lambda_{ij}$ have to fulfill the ``normalization''
\begin{equation}
\label{eq:norm_lambda}
\int \mathrm{d}\tilde{\VEC{\alpha}} \, \lambda_{ij} = \overline{\lambda_{ij}} = 0
\end{equation}
to guarantee overall normalization of $p$ in \eqref{eq:p}.

Analogously to the above procedure for \eqref{eq:hierarchy1}, we
next analyse the solvability condition for the second order
equation \eqref{eq:hierarchy2}. Plugging in the results we derived
so far, namely \eqref{eq:p0}, \eqref{eq:FPconvective} and \eqref{eq:p1},
we obtain in a straightforward way
\begin{equation}
\label{eq:FPdiffusive}
\frac{\partial\rho^{(0)}}{\partial\tau}
- D^{\mathrm{eff}}_{ij} \frac{\partial}{\partial X_i}\frac{\partial}{\partial X_j} \rho^{(0)}
= 0 \, ,
\end{equation}
with the effective diffusion coefficient on large scales
\begin{equation}
\label{eq:Deff}
D^{\mathrm{eff}}_{ij}
= \kB T \, (\overline{\gamma^{-1}})_{ij} - \overline{(\gamma^{-1})_{ik}\lambda_{jl}} \, f_k f_l \, ,
\end{equation}
where we used \eqref{eq:norm_lambda} to arrive at the given form of the second term.

As anticipated when introducing the scaling ansatz \eqref{eq:scales},
the main results \eqref{eq:FPconvective} and
\eqref{eq:FPdiffusive} of the multi-scale
analysis consist in an equation,
which describes convective motion with an effective velocity 
on time scales $\vartheta$ and large spatial scales
(eq.~\eqref{eq:FPconvective}),
and an equation, which governs diffusion
with an effective diffusion coefficient
on time scales $\tau$ and large spatial scales (eq.~\eqref{eq:FPdiffusive}).
In order to combine them into one effective equation
of motion,
we switch back to the original variables $t$ and $\VEC{x}$,
and use that the marginal density $\rho(t,\VEC{x})$ in translation space only
is obtained by integrating $p$ over orientational degrees of freedom,
$\rho=\int \mathrm{d}\VEC{\alpha}\,p$.
From \eqref{eq:p} we conclude that $\rho = \rho^{(0)}$ in lowest order $\EPS$.
Using \eqref{eq:derivatives} (and $\partial\rho^{(0)}/\partial\theta=0$)
we find $\partial\rho/\partial t = \EPS\,\partial\rho^{(0)}/\partial\vartheta + \EPS^2\partial\rho^{(0)}/\partial\tau$
in lowest order.
Plugging in our results \eqref{eq:FPconvective} and \eqref{eq:FPdiffusive},
and noticing that $\VEC{x}=\EPS^{-1}\VEC{X}$
(see \eqref{eq:scales}) we finally arrive at
\begin{equation}
\label{eq:FPeff}
\frac{\partial\rho}{\partial t}
+ \frac{\partial}{\partial x_i} \left(
V_i - D^{\mathrm{eff}}_{ij} \frac{\partial}{\partial x_j}
\right) \rho = 0 \, .
\end{equation}
Although it is given in the original variables $t$ and $\VEC{x}$, we know
from the way it has been derived that this effective forward Fokker-Planck
equation is a valid description of the dynamics of our system \eqref{eq:balance}
only in the long-term regime $t\gg\tau_c$ and $|\VEC{x}| \gg l_c$.
As usual in multi-scale schemes \cite{pavliotis08}, the explicit expressions
for the ``effective coefficients''
$V_i$ and $D^{\mathrm{eff}}_{ij}$ are obtained
from solving an auxiliary equation on the small scales,
here eq.~\eqref{eq:FP_lambda}, and from averaging over these small
scales (see eqs.~\eqref{eq:V} and \eqref{eq:Deff}).

\section{Calculation of the effective coefficients}
The relevant averages over small scale rotational variables
in \eqref{eq:V}, \eqref{eq:FP_lambda} and \eqref{eq:Deff}
are of the form $\overline{(\gamma^{-1})_{ij}}$ and
$\overline{(\gamma^{-1})_{ik}\lambda_{jl}}$.
Since these averages are performed over a uniform distribution
of orientations, the resulting tensors should be invariant under rotation.
We can thus use invariance theory to show that
\begin{subequations}
\label{eq:av}
\begin{eqnarray}
\overline{(\gamma^{-1})_{ij}} & = & \frac{1}{3}\Tr(\gamma^{-1})\delta_{ij}
\label{eq:avTensor}
\, , \\
\overline{(\gamma^{-1})_{ik}\lambda_{jl}} & = &
\left[ -\frac{1}{15}\delta_{ik}\delta_{jl} + \frac{1}{10} ( \delta_{ij}\delta_{kl} + \delta_{il}\delta_{jk} ) \right]
\nonumber\\
&& \qquad
\times \Tr(\gamma^{-1}\lambda)
\label{eq:avTensorTensor}
\, .
\end{eqnarray}
\end{subequations}
In analogy to \eqref{eq:avTensor}
we also find $\overline{\lambda_{ij}}=\delta_{ij}\Tr(\lambda)/3$,
so that $\lambda$ has to be traceless according to \eqref{eq:norm_lambda}.
The property $\Tr(\lambda)=0$ has already been used
to simplify \eqref{eq:avTensorTensor}.

In order to find the solution to the auxiliary equation \eqref{eq:FP_lambda},
we first rewrite its right-hand inhomogeneity using \eqref{eq:avTensor},
\begin{equation}
\label{eq:gammahat}
(\hat\gamma^{-1})_{ij} = (\gamma^{-1})_{ij}-\frac{1}{3}\Tr(\gamma^{-1})\delta_{ij}
\, ,
\end{equation}
i.e.\ $(\hat\gamma^{-1})_{ij}$ denotes the traceless part 
of the inverse friction tensor.
This tensor $(\hat\gamma^{-1})_{ij}$ depends on the particle
orientation via rotation matrices $\TENSOR{R}=\TENSOR{R}(\VEC{\alpha})$, i.e.\
$(\hat\gamma^{-1})_{ij} = (\TENSOR{R}\hat\Gamma^{-1}\TENSOR{R}\TRANS)_{ij}$,
where the traceless inverse friction tensor $\hat\Gamma^{-1}$ represents
an arbitrary reference configuration and thus is constant. Typically, it
is chosen such that the principal axes of the ellipsoid are
aligned with the coordinate axes of the laboratory frame,
because then $\hat\Gamma_{ij}$ is diagonal.
Assuming that the mobility tensor $\gamma^{-1}$ and the rotational
diffusion tensor $\TENSOR{Q}$ appearing in $\OP{M}^\dagger$
can be diagonalized in the same frame,
this is also the configuration of the ellipsoid in which $\TENSOR{Q}$
(with components $Q_{ij}$) is diagonal.
We may therefore expect that even $\lambda_{ij}$ ``inherits'' that
property such that it can be written as
$\lambda_{ij} = (\TENSOR{R}\Lambda\TENSOR{R}\TRANS)_{ij}$
with $\Lambda$ being diagonal and constant.

To find the explicit form of the rotation matrices $\TENSOR{R}(\VEC{\alpha})$
and, in particular, the rotary diffusion operator $\OP{M}^\dagger$
for a specific parametrization of orientation may be quite
cumbersome. However, in \cite{rallison78} Rallison showed that
we can write
\begin{equation}
\OP{M}^\dagger = \epsilon_{ikl}R_{km}\frac{\partial}{\partial R_{lm}} Q_{ij} \epsilon_{jpq}R_{pn}\frac{\partial}{\partial R_{qn}}
\, ,
\end{equation}
independent of the specific representation of rotation, as
only derivatives $\partial/\partial R_{ij}$ with respect to
the ${ij}$-component of the rotation matrix appear
($\epsilon_{ijk}$ is the Levi-Civita tensor).
With this expression and the above ansatz for $(\hat\gamma^{-1})_{ij}$
and $\lambda_{ij}$, \eqref{eq:FP_lambda} reduces to an algebraic
equation, which is easily inverted. After some lengthy but
straightforward algebra, we find the solution
\begin{subequations}
\begin{equation}
\label{eq:lambda}
\lambda_{ij} = \frac{1}{6\Delta} \left[ \Tr(\TENSOR{Q}\hat\gamma^{-1})\delta_{ij} - 3(\TENSOR{Q}\hat\gamma^{-1})_{ij} \right]
\, ,
\end{equation}
with
\begin{equation}
\Delta = Q^{(1)}Q^{(2)} + Q^{(2)}Q^{(3)} + Q^{(3)}Q^{(1)}
\, ,
\end{equation}
\end{subequations}
and the eigenvalues $Q^{(i)}$ of the rotary diffusion tensor $\TENSOR{Q}$.

Plugging the expressions \eqref{eq:av} and \eqref{eq:lambda} into \eqref{eq:V}
and \eqref{eq:Deff}, we obtain the final results for the effective coefficients,
\begin{subequations}
\label{eq:final}
\begin{eqnarray}
V_i & = & \frac{1}{3}\Tr(\gamma^{-1}) f_i
\, ,
\label{eq:V_final} \\
D^{\mathrm{eff}}_{ij} & = & \bar{D}\,\delta_{ij} + \kappa_{ij}
\, ,
\label{eq:Deff_final}
\end{eqnarray}
where we re-used the abbreviation $\bar{D} = \Tr(\TENSOR{D})/3 = \kB T \Tr(\gamma^{-1})/3$
in \eqref{eq:Deff_final} and defined
\begin{equation}
\label{eq:kappa_final}
\kappa_{ij} = \frac{1}{2\Delta} \Tr\left( \hat\gamma^{-1}\TENSOR{Q}\hat\gamma^{-1} \right)
\left( \frac{1}{30}f_i f_j + \frac{1}{10} \VEC{f}^2 \delta_{ij} \right)
\, .
\end{equation}
\end{subequations}
For the specific choice $\VEC{f}=(f,0,0)$, the effective diffusion tensor becomes
diagonal with
\begin{equation}
\label{eq:kappa_ii}
\kappa_{11} = \frac{1}{15\Delta} \Tr\left( \hat\gamma^{-1}\TENSOR{Q}\hat\gamma^{-1} \right) f^2
\, , \quad
\kappa_{22}=\kappa_{33} = \frac{3}{4}\kappa_{11} \, .
\end{equation}
The $11$-component quantifies effective diffusion parallel,
and the $22$- and $33$-component perpendicular
to the direction of the external force $\VEC{f}$.

\section{Numerical simulations}
\begin{figure}
\onefigure[width=0.8\columnwidth]{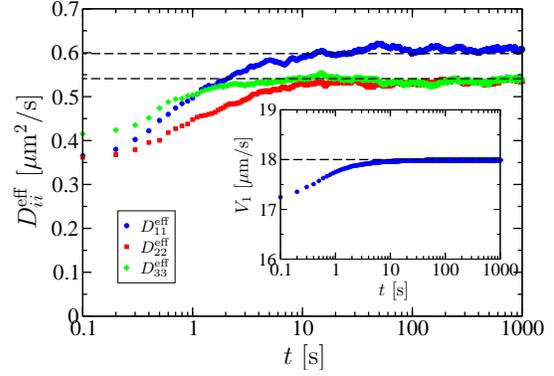}
\caption{Comparison of numerical simulations with the theoretical predictions
\eqref{eq:final}. The symbols show the three different components of
the net diffusion coefficients $D^{\mathrm{eff}}_{ii}$ and the net
translational velocity $V_1$ (inset)
as a function of time,
averaged over 10000
realization of the Gaussian noise source, but for identical initial position
and orientation of the ellipsoid.
The dashed horizontal lines represent the theoretical long-term
predictions from \eqref{eq:final},
$D^{\mathrm{eff}}_{11} = \unit[0.60]{\mu m^2/s}$,
$D^{\mathrm{eff}}_{22} = D^{\mathrm{eff}}_{33} = \unit[0.54]{\mu m^2/s}$
and $V_1 = \unit[0.18]{\mu m/s}$
(the other net velocity components are zero, not shown).
For comparison, the
purely thermal contribution to the net diffusion is $\bar{D}=\unit[0.37]{\mu m^2/s}$.
The simulated ellipsoid has semi-axis lengths
\unit[0.3]{$\mu$m}, \unit[0.6]{$\mu$m} and \unit[0.9]{$\mu$m}.
In water, its translational friction coefficients
for moving in the direction of these axes are
(to two digits precision)
\unit[12]{fN\,s/$\mu$m},
\unit[11]{fN\,s/$\mu$m}, and
\unit[10]{fN\,s/$\mu$m}, respectively,
and its friction coefficients for rotation around these axes are
\unit[7.1]{fN\,$\mu$m\,s},
\unit[6.8]{fN\,$\mu$m\,s}, and
\unit[4.1]{fN\,$\mu$m\,s} \cite{brenner67}.
The thermal energy is set to $\kB T =\unit[4.1]{fN\mu m}$, corresponding to
room temperature (\unit[300]{K}).
The external force is $\VEC{f}=(\unit[200]{fN},\unit[0]{fN},\unit[0]{fN})$.}
\label{fig:1}
\end{figure}

We compare our main results \eqref{eq:final} to numerical simulations
of the original equations of motion \eqref{eq:balance} 
for a representative ellipsoid with ratio $1:2:3$ between its
three semi-axes.
For the simulations, we choose quaternions as a concrete
parametrization of rotation, i.e.\ the model \eqref{eq:balance}
is supplemented by the equation of motion for the
quaternion $q$,
\begin{equation}
\label{eq:q}
\dot{q} = \frac{1}{2} \omega \circ q \, ,
\end{equation}
where the symbol $\circ$ denotes a \emph{quaternion} product
\cite{coutsias04}
evaluated in the Stratonovich sense \cite{mazo02,gardiner83,vankampen87},
and where $\omega$
is the angular velocity in the laboratory frame from \eqref{eq:Frot},
represented as a pure quaternion \cite{coutsias04}.
We solve \eqref{eq:balance} and \eqref{eq:q} using the
Euler algorithm with time-step \unit[0.1]{ms}.
The explicit values for the
(eigenvalues of the) friction tensors $\gamma$ and $\eta$
of the ellipsoid
are calculated from the exact expressions
given in \cite{brenner67}.
For the simulations we choose an
ellipsoid with semi-axis lengths
\unit[0.3]{$\mu$m}, \unit[0.6]{$\mu$m} and \unit[0.9]{$\mu$m}.
The results of the simulations averaged over
10000 realizations of the noise sources
(with identical initial conditions)
are shown in Fig.~\ref{fig:1}.
The crossover from short-term to long-term diffusion is
clearly visible to occur at a time around the order of a second,
well comparable to the estimation $\tau_c=\unit[1.4]{s}$
from \eqref{eq:tau_c}.
The long-term coefficients obtained from these
simulations are in perfect agreement with the theoretical
predictions \eqref{eq:final}.

\section{Discussion and Conclusions}
The effective diffusion coefficient \eqref{eq:Deff_final}
is composed of two contributions.
The first one, $\bar{D}\delta_{ij}$, is the purely thermal
diffusion of the center of the ellipsoid, averaged over all
particle orientations. The second one, $\kappa_{ij}$, is a
dispersion effect stemming from the variations of the translational
velocity with diffusive changes of the particle orientation
relative to the direction of the external force $\VEC{f}$.
This contribution is anisotropic, the dispersion effect
is stronger in the direction of the force than perpendicular
to it. Remarkably, the ratio $4/3$ between the
parallel and perpendicular component (see \eqref{eq:kappa_ii})
is completely independent of the specific shape of the ellipsoid.
Furthermore, this anisotropy is only present in three dimensions.
For two dimension, the effective diffusion tensor
analogous to \eqref{eq:kappa_final} turnes out to be \emph{isotropic};
it has been obtained in \cite{grima07} by formally solving the Langevin
equations of motion for translation and rotation in the plane.


It is straightforward to verify that \eqref{eq:Deff_final}
and \eqref{eq:kappa_ii} agree with the results obtained by
Brenner in \cite{brenner81}. 
However, unlike \cite{brenner81} we have here shown that 
\eqref{eq:FPeff} and \eqref{eq:final} follow from a
rigorous perturbative multi-scale technique which has been 
widely used in physics and applied mathematics over the last 10-20 years.
Mean convective velocity and effective diffusion tensor
are both computed by solving an auxiliary equation
(see eq.~\eqref{eq:FP_lambda}),
and by performing averages over
the stationary rotational distribution of the original model
(see eqs.~\eqref{eq:av}).
In the case studied here, where there is no torque,
this procedure is pretty straightforward:
the stationary rotational distribution
is uniform, and the solution of the auxiliary equation
can be obtained by a simple ansatz.
In more general settings, the stationary distribution as well as
the solution of the auxiliary equation would have to be computed
numerically, in analogy to the description of scalar transport in 
compressible flows presented in~\cite{vergassola97}; we will leave
this for future work.
Specifically, such more general settings
may include external torques \cite{guell10,sandoval13},
or other types of external ``forces'', like hydrodynamic flows
or phoretic mechanisms which drive particle motion.
We finally remark that experiments for
measuring the long-term convective
and diffusive motion of ellipsoidal particles
in three dimensions, and for
verifying our theoretical
predictions in \eqref{eq:final},
could be performed along the lines
of \cite{fung13}.

\acknowledgments
Financial support
by the Swedish Science Council (Vetenskapsr{\aa}det)
under the grants 621-2012-2982 and 621-2013-3956
is acknowledged.

\end{document}